\documentclass[acmsmall,screen,nonacm]{acmart}

\graphicspath{{figures/}}

\usepackage{xspace}
\usepackage{tcolorbox}
\usepackage{tabularx}
\usepackage{tikz}
\usetikzlibrary{positioning,arrows.meta,calc,fit}
\usepackage{fontawesome5}

\setcopyright{none}

\newcommand{\approach}{BBCFixer\xspace}
\newcommand{\benchmark}{BBCBench\xspace}

\begin{document}

\title{Who Broke Me? Execution-Guided Repair of Behavioral Dependency Breaks}

\author{Juntao Chen}
\orcid{0009-0006-7815-540X}
\affiliation{%
  \institution{The University of Melbourne}
  \city{Melbourne}
  \country{Australia}}
\email{juntao.chen@student.unimelb.edu.au}

\author{Tingting Bi}
\orcid{0000-0003-2748-1249}
\authornote{Corresponding author.}
\affiliation{%
  \institution{The University of Melbourne}
  \city{Melbourne}
  \country{Australia}}
\email{tingting.bi@unimelb.edu.au}

\author{Yanlin Wang}
\orcid{0000-0001-7761-7269}
\affiliation{%
  \institution{Sun Yat-sen University}
  \city{Guangzhou}
  \country{China}}
\email{wangylin36@mail.sysu.edu.cn}

\author{Patanamon Thongtanunam}
\orcid{0000-0001-6328-8839}
\affiliation{%
  \institution{The University of Melbourne}
  \city{Melbourne}
  \country{Australia}}
\email{patanamon.t@unimelb.edu.au}

\renewcommand{\shortauthors}{Chen et al.}

\begin{abstract}
Dependency upgrades can break downstream projects without changing the library interface. Such \textbf{behavioral breaking changes} are difficult for developers to fix, because the failing test does not always point to the \emph{root API}, the upstream API that causes the break. Existing LLM-based repair methods obtain evidence for the repair from compiler feedback or library documentation. However, a behavioral break produces no compiler feedback and is often undocumented. Agents that receive no upgrade evidence also usually do not retrieve upstream evidence themselves, and most of their failed repairs do not identify the root API.

The library diff provides useful upstream evidence, but the root API must first be identified to select the relevant part of the diff. We present \approach{}, a repair method that runs the failing test under the old and new library versions, ranks the calls whose return value differs to identify a candidate root API, and filters the library diff with that candidate. To evaluate repairs of behavioral breaks, we further introduce \benchmark{}, a benchmark of 100 behavioral dependency breaks in Python and JavaScript.

We compare \approach{} on \benchmark{} with two baselines: (1) Pure Agent, an agent that receives no upgrade evidence, and (2) BDUpdater, a method that mines library documentation. \approach{} raises the average pass rate by up to 17\% relative to Pure Agent and needs up to 27\% fewer agent steps per successful repair. \approach{} thus provides a way to select the upstream evidence behind a behavioral break, so that an LLM agent can repair behavioral breaking changes more effectively and efficiently.

\end{abstract}



\maketitle

\section{Introduction}
\label{sec:introduction}

Modern software depends heavily on third-party libraries, whose upgrades deliver security fixes, bug fixes, and new functionality. Yet developers often postpone these upgrades: 81.5\% of the systems studied by Kula et al.~\cite{10.1007/s10664-017-9521-5} retain outdated dependencies. One reason is that a new library version can introduce breaking changes that cause existing downstream code to fail~\cite{chen2026breakingchangessoftwareecosystems,10.1007/s10664-014-9325-9}.

Breaking changes are commonly categorized as \textbf{syntactic} or \textbf{behavioral}~\cite{chen2026breakingchangessoftwareecosystems,10.1145/3643782}. A syntactic breaking change alters the library interface, for example by removing an API or changing its signature. Such changes produce compiler or import errors that identify the affected symbol and direct developers toward the upgrade~\cite{10.1145/3092703.3092721}. A behavioral breaking change is more implicit and harder to diagnose: the interface remains intact, but the \textit{run-time behavior changes}. The downstream project still builds, but the break instead appears as a failing test whose error may not point to the library or API that causes the break. Such breaks are difficult for developers to fix in practice. In the real-world bug reports studied by Mostafa et al., 18\% of the client bugs caused by behavioral backward incompatibilities remain unfixed, although most developers try to fix them~\cite{10.1145/3092703.3092721}. Therefore, this paper focuses on the repair of these behavioral dependency breaks.

Recent work repairs breaking changes either with single-call LLM pipelines~\cite{10.1007/s10664-026-10835-1,10.1145/3803437.3806404,yang2026deprepairllmbasedsourcecoderepair} or with LLM agents~\cite{10.1145/3729366,reyes2026agenticgenerationasttransformation,tawosi2025llmagentsautomateddependency,10.1145/3808116}. Both pipelines and agents need \emph{evidence} that connects the downstream failure to the upstream change, and obtain it from compiler feedback or library documentation. For example, Byam prompts the LLM with the compilation errors of the broken build~\cite{10.1007/s10664-026-10835-1}, and BDUpdater mines breaking change records from the upstream repository~\cite{10.1145/3808116}. Neither source works for a behavioral break. Compiler feedback is absent in behavioral breaks, and documentation is unreliable: Mostafa et al.\ find that only 27.7\% of the behavioral backward incompatibilities they detect are documented~\cite{10.1145/3092703.3092721}.

\textbf{Challenge 1}: in repairing breaking changes with LLM agents, \textbf{cost matters in practice}. One library update breaks many downstream projects, and each must be repaired separately~\cite{reyes2026agenticgenerationasttransformation}. At this scale, the cost of each agent run dominates the total cost~\cite{kapoor2024aiagentsmatter}, so low-priced models are the practical choice. Higher-priced models also do not consistently give better cost-efficiency in LLM-based program repair~\cite{liu2026rethinkingautomatedprogramrepair}. Mining documentation adds to this cost. In our experiments (Section~\ref{sec:rq2}), BDUpdater raises the cost of each successful repair by up to 132\%. The search for documentation consumes the agent steps that the repair itself needs. As a result, recent work either lacks the evidence that a behavioral break requires, or searches for it at a high cost.

\textbf{Challenge 2} is not merely generating a patch, but\textbf{ selecting the evidence that makes the required adaptation apparent}. This evidence must be condensed before the agent begins, rather than mined by the agent during the run. The failure is observed in a downstream test, whereas its explanation resides in the upstream version history. We call this separation the \emph{evidence-selection gap}. In our Pure Agent setting (Section~\ref{sec:setup}), the agent receives no upgrade evidence but can inspect the repository and the installed library. However, most agent runs consult neither textual nor code-level information about the upgraded library, and failure to identify the relevant upstream behavioral change is the dominant failure pattern (Section~\ref{sec:rq1}).

The library diff provides a direct source of upstream evidence but requires selection, because the full diff is far too large for an agent to read (Section~\ref{sec:motivation_upgrade_evidence}). The diff covers changes to many library APIs, and a failing downstream test often reaches several of them. However, the break is caused by the behavior change of one of them, the \emph{root API}. The root API is hard to identify, because the failing test makes many calls into the library and the error message often does not name the API. Our key insight is therefore to use differential execution as a selector for \textit{upstream evidence}. Differential execution runs the same failing test under the old and new library versions and reveals the library calls whose return value differs. Prior work uses differential execution to detect incompatibilities~\cite{9350237,10.1145/3582569,monce2025gilesi,10.1145/3744916.3764527}, whereas we use it to select evidence for repair. We propose \approach{}, which ranks these diverging calls, takes the highest-ranked one as the \emph{first divergence}, and uses its API as a \emph{candidate root API}. \approach{} then filters the library diff with that candidate and gives the repair agent the first divergence and the filtered library diff as evidence. \approach{} turns a large upstream version change into compact, downstream-specific repair evidence before the agent begins its search. In this way, the agent can spend its steps on the repair itself and keep its context small. \approach{} therefore lowers the cost of each successful repair compared with mining documentation.

Furthermore, evaluating repairs of behavioral breaks also requires a suitable benchmark. Existing dependency-upgrade benchmarks contain few confirmed behavioral breaks (Section~\ref{sec:benchmark_motivation}), and most of them also lack either reference fixes or a public harness that reproduces each case. A reference fix serves as the oracle: its test changes encode the correct new behavior, so they can check an agent's repair. We therefore construct \benchmark{} to address these problems. \benchmark{} contains 100 behavioral dependency breaks in real Python and JavaScript projects, each with a reference fix and a reproducible harness for the repair task.

We then compare \approach{} with two baselines (i.e., Pure Agent and BDUpdater) (Section~\ref{sec:setup}) on \benchmark{}, using two budget models, qwen3-coder and GLM-4.5-Air, over three rounds. We run three rounds because the outcome of a single run varies between repetitions. \approach{} raises the pass rate above both settings under both models in every round. \approach{} raises the average pass rate by up to \textbf{17}\% over Pure Agent and by up to \textbf{53}\% over BDUpdater. \approach{} also needs fewer agent steps per successful repair: up to \textbf{27}\% fewer than Pure Agent and up to \textbf{58}\% fewer than BDUpdater.

In summary, this paper makes the following contributions:

\textbf{Problem formulation and empirical finding}. We characterize the repair of behavioral breaking changes as an evidence-selection problem. Our trajectory analysis shows that repository access alone does not lead the agent to retrieve the necessary upstream evidence and most failing runs never identify the root API.

\textbf{Approach}. We propose \approach{}, a repair method that uses differential execution to rank the calls whose return value differs, identify a candidate root API, and filter the library diff with that candidate. The first divergence and the filtered library diff then serve as repair evidence for the agent.

\textbf{Benchmark}. We construct \benchmark{}, a benchmark of 100 confirmed behavioral dependency breaks in real Python and JavaScript projects. Each case includes a reproducible harness and a reference fix whose test changes encode the intended post-upgrade behavior.

\textbf{Evaluation and implications}. We evaluate \approach{} on \benchmark{} against Pure Agent and BDUpdater. \approach{} achieves a higher average pass rate under both models, with fewer agent steps and lower cost per successful repair. The gain of \approach{} depends on how the evidence is built and concentrates on the cases whose evidence points to the labeled root API. The evaluation indicates that building failure-specific evidence is a useful complement to improving the agent itself.

\section{Background and Motivation}
\label{sec:background}

In this section, we discuss the background and the motivation of this study. Section~\ref{sec:bbc_definition} defines behavioral breaking changes and the repair task. Section~\ref{sec:motivation_upgrade_evidence} shows that the library diff of the upgrade carries the evidence but is too large to use without filtering.

\subsection{Behavioral Breaking Changes and the Repair Task}
\label{sec:bbc_definition}

\begin{figure}[t]
  \centering
  \includegraphics[width=\linewidth]{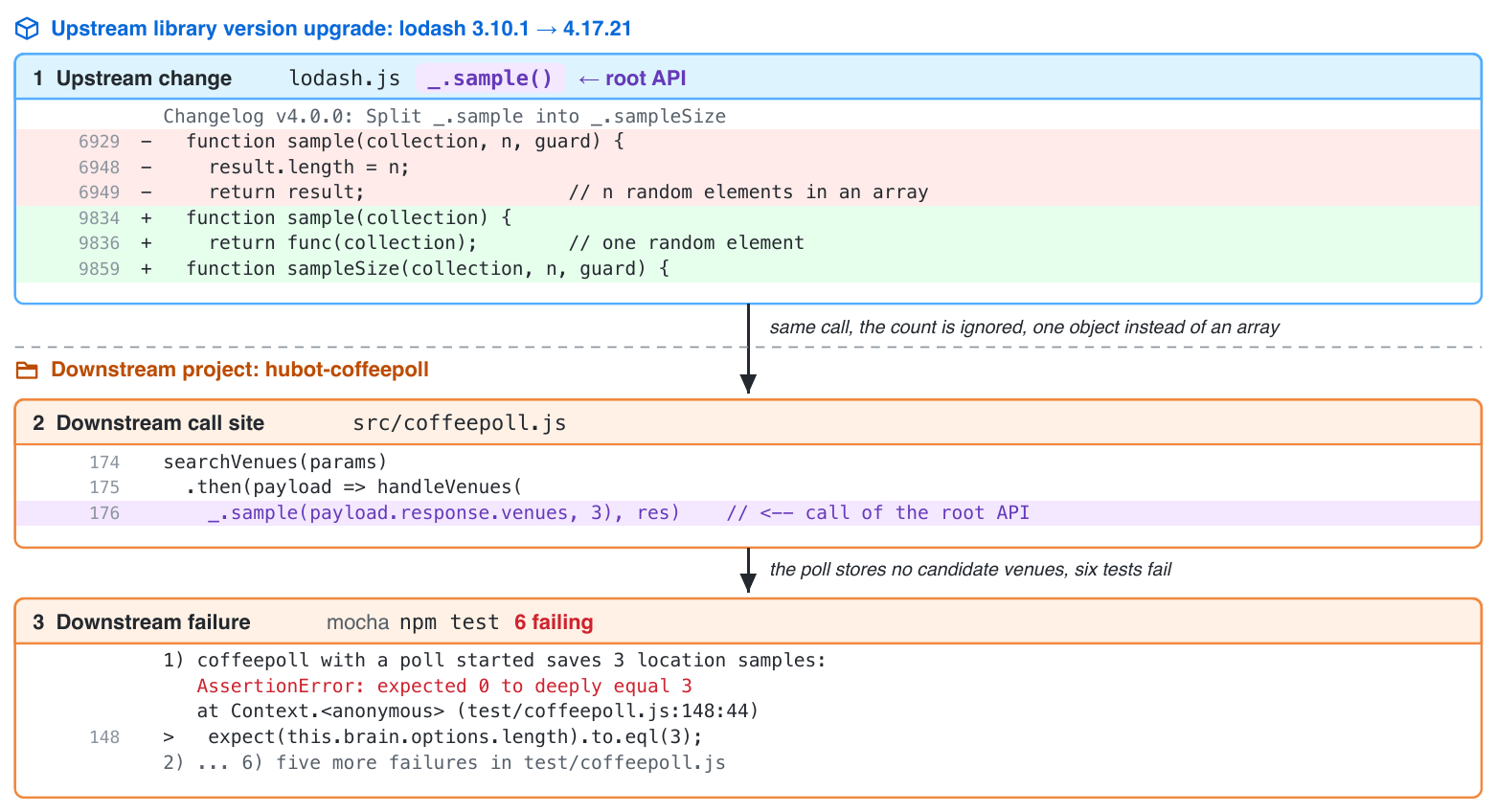}
  \caption{An example of a behavioral breaking change and its manifestation in a downstream project.}
  \Description{Three stacked panels in the style of GitHub and mocha, with
  an upstream lane on top and a downstream lane below, separated by a dashed
  line. Panel 1, the upstream change in lodash, marks \_.sample() as the root
  API and shows its diff between 3.10.1 and 4.17.21: the old version takes a
  count n and returns n random elements in an array, the new version takes
  only the collection and returns one random element, and a new function
  \_.sampleSize(collection, n) is added. Panel 2, the downstream call site in
  hubot-coffeepoll, shows the unchanged call \_.sample(payload.response.venues,
  3), marked as the call of the root API. Panel 3, the downstream failure,
  shows mocha reporting 6 failing tests, the first with AssertionError:
  expected 0 to deeply equal 3 at test/coffeepoll.js line 148.}
  \label{fig:bbc-consequence}
\end{figure}

A behavioral breaking change leaves the interface intact but alters the library behavior, such as the type of a return value or a default setting. Figure~\ref{fig:bbc-consequence} shows an example, in which the downstream project hubot-coffeepoll, a chat-bot plugin for coffee-shop polls, upgrades lodash from 3.10.1 to 4.17.21. In 3.10.1, \texttt{\_.sample()} called with a count of 3 returns an array of three random elements. In 4.17.21, \texttt{\_.sample()} ignores the count and returns a single element, and the new function \texttt{\_.sampleSize()} returns the three elements instead. The downstream call still runs, since JavaScript does not check the number of arguments. The plugin, however, now receives one object instead of an array, stores an empty list of candidate venues, and fails six tests. The case shows two properties of behavioral breaking changes. First, the change is hidden, since the failing assertions and their stack traces name only the test file, not lodash. Second, the repair needs information that the failure does not carry: the developer must find that \texttt{\_.sample()} no longer returns an array and that \texttt{\_.sampleSize()} now does, before rewriting the call.

The repair task for the agent is defined as follows. The agent receives a downstream repository that breaks after one dependency is upgraded to a target version \texttt{v\_new}. The agent modifies the downstream source code so that the repository builds and passes its tests on \texttt{v\_new}. Downgrading the dependency, removing it, or switching to another version is not allowed. Section~\ref{sec:setup} lists the full constraints and budgets.

\subsection{Filtering the Library Diff}
\label{sec:motivation_upgrade_evidence}

Because documentation often leaves a behavioral change unrecorded, the library diff provides an alternative source of evidence that comes from the two library versions. The diff records how the implementation changed and may also contain revised library tests that specify the new behavior.

\begin{figure}[t]
  \centering
  \includegraphics[width=\linewidth]{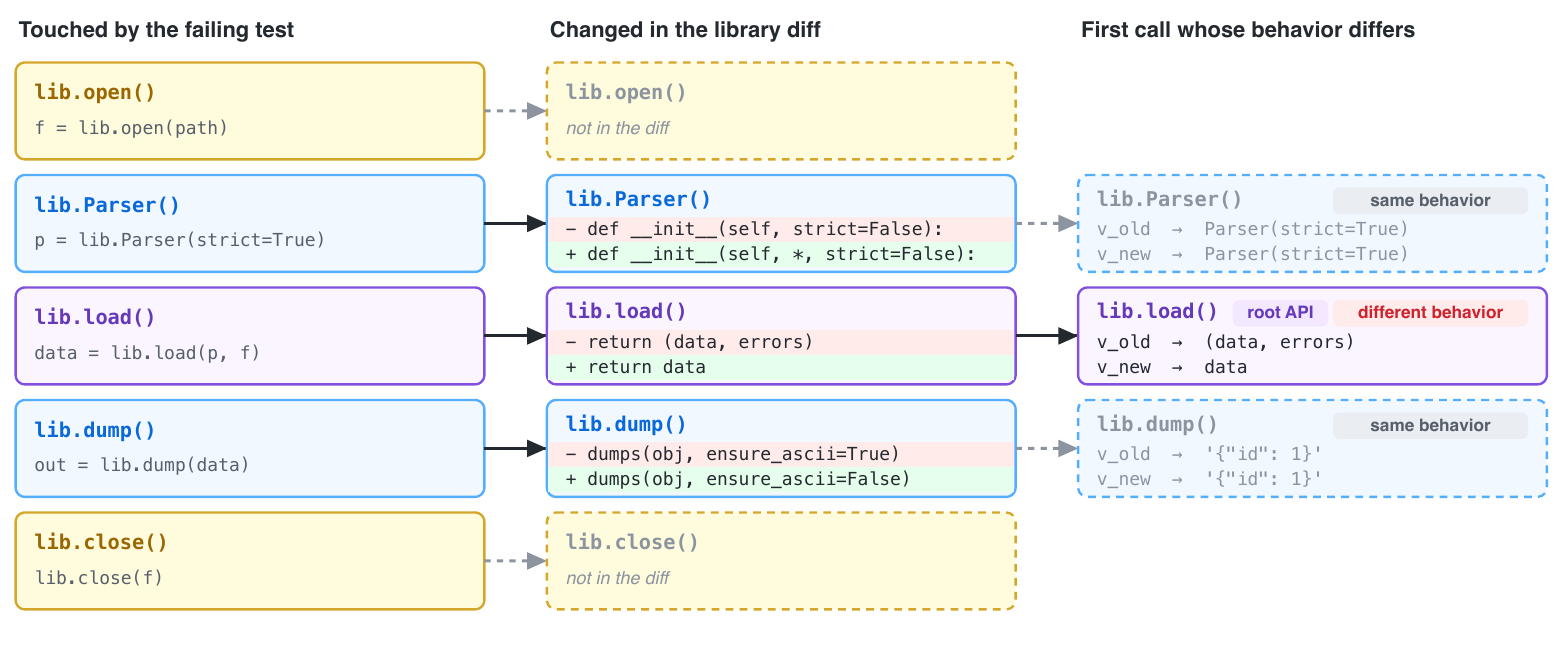}
  \caption{Narrowing the upstream APIs touched by one failing test to the root API.}
  \Description{Three columns of API cards aligned in call order. Column 1
  lists the five APIs that the failing test touches. Column 2 keeps the
  three that the library diff changes and shows a two-line diff for each.
  Column 3 shows the return value under the old and the new version, two
  candidates return the same value, and lib.load() is marked as the root
  API because its return value changes.}
  \label{fig:root-api-funnel}
\end{figure}

The library diff, however, is far too large for an agent to read. Across the 100 benchmark cases (Section~\ref{sec:benchmark}), the diff between the two library versions has a median of 115{,}866 lines, spread over a median of 168 files. Feeding the whole diff to the agent buries the few relevant lines, so the diff must be filtered down to the fragment that explains the break. Effective filtering keeps the root API, and identifying that API is difficult. One failing test touches several upstream APIs, the interface stays unchanged, and the error message names none of them. Figure~\ref{fig:root-api-funnel} shows how the candidates narrow on an illustrative failing test. Across the benchmark cases, one failing test touches a median of 5 distinct upstream APIs, of which the library diff changes a median of 3. Keeping only the changed APIs therefore still leaves several candidates, and selecting among them needs evidence that depends on how the downstream actually uses the library. In the figure, only \texttt{lib.load()} returns a different value under the two versions, and this difference singles it out as the root API.

\section{Research Design}
\label{sec:research_design}

Our research design connects three components. First, \approach{} transforms the behavioral difference observed in a downstream test into compact upstream repair evidence (Section~\ref{sec:approach}). Second, \benchmark{} provides 100 behavioral dependency breaks in real downstream projects, each with reproducible intact, broken, and fixed states and an automated evaluation script (Section~\ref{sec:benchmark}). Third, the experimental setup evaluates the complete repair workflow and its evidence configurations under two models (Section~\ref{sec:setup}).

\subsection{Our Approach: \approach{}}
\label{sec:approach}

\begin{figure}[t]
  \centering
  \includegraphics[width=\linewidth]{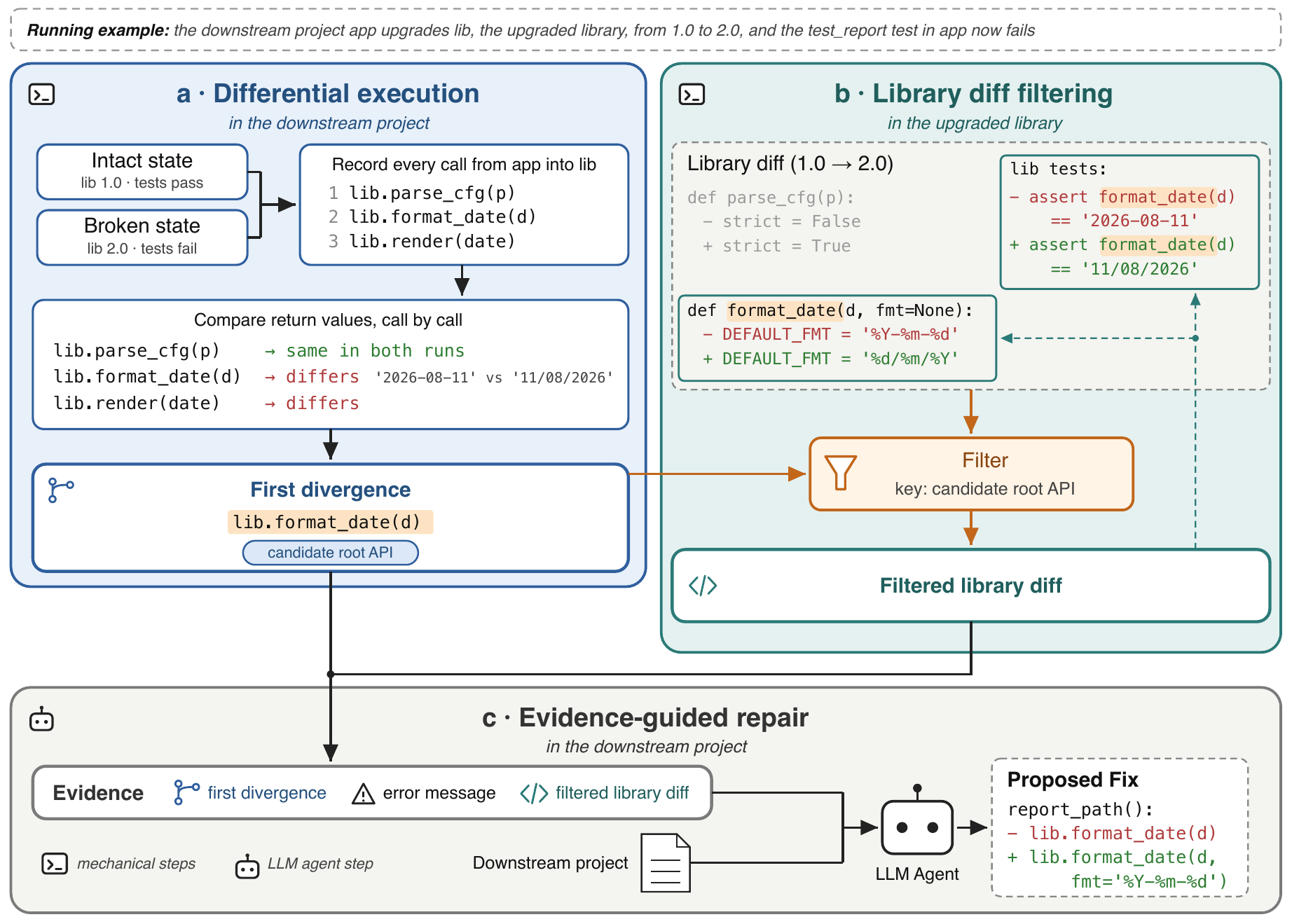}
  \caption{Overview of \approach{}.}
  \Description{Overview of \approach{}.}
  \label{fig:approach}
\end{figure}

\approach{} addresses the evidence-selection gap identified in Section~\ref{sec:introduction}. Its goal is not to ask the repair agent to inspect the entire upstream version history, but to construct a compact explanation of the observed downstream failure before repair begins. As Figure~\ref{fig:approach} shows, the pipeline has three stages. It first compares the failing test under the intact and broken dependency versions to prioritize a candidate upstream API (Section~\ref{sec:approach_diffexec}). It then uses this candidate and the failure symptoms to extract relevant fragments from the library diff (Section~\ref{sec:approach_filter}). Finally, it presents the first divergence and the filtered library diff to the repair agent as failure-specific evidence (Section~\ref{sec:approach_loop}).

\subsubsection{Differential Execution}
\label{sec:approach_diffexec}

\textit{\textbf{Panel a}} of Figure~\ref{fig:approach} shows this step. \approach{} runs the failing test twice, once in the intact state with the old library version and once in the broken state with the upgraded version. During each run, \approach{} records every call that enters the upgraded library from outside it, with its arguments and its return value. It then aligns the recorded calls across the two executions and compares their observed return values. A call is a \emph{divergence} when its return value differs between the old and new library versions. Multiple calls may diverge during one test execution, so \approach{} ranks these divergences to select the one most likely to be relevant to the downstream failure. Calls originating in downstream code rank above calls originating in other libraries. Among downstream calls, those from files appearing in the failing stack trace rank higher; remaining ties are resolved by execution order. We call the highest-ranked divergence the \emph{first divergence} and use its API as a candidate root API.

Figure~\ref{fig:approach} also carries a running example. The failing test first calls \texttt{parse\_cfg}, which returns the same value in both executions. The two later calls, \texttt{format\_date} and \texttt{render}, both diverge. The call to \texttt{format\_date} returns \texttt{'2026-08-11'} with the old library version and \texttt{'11/08/2026'} with the new version, and \texttt{render} then returns a different value because it receives this changed date string as input. Both diverging calls originate in the same downstream file, so the ranking resolves by execution order and \approach{} selects the earlier divergence, \texttt{format\_date}, as the candidate root API.

\subsubsection{Library Diff Filtering}
\label{sec:approach_filter}

\textit{\textbf{Panel b}} of Figure~\ref{fig:approach} shows this step. The library diff includes changes to both library source files and library tests. \approach{} next reduces this diff to repair-relevant fragments, using the candidate root API as the primary filter key. \approach{} also extracts symptom symbols from the broken-state error message, including exception names, failed assertion values, and library frames in the stack trace, and adds them as further filter keys. When differential execution reports no divergence, the primary key falls back to the upstream symbols that the downstream code uses.

The filtering keeps changed lines containing a filter key, each with a few lines of surrounding context. In the running example, filtering with the key \texttt{format\_date} keeps two fragments: the changed default-format implementation, and the revised library assertion that specifies the new expected output. The resulting filtered library diff provides both an implementation-level explanation of the change and, when available, an executable specification of the new behavior.

\subsubsection{Evidence-Guided Repair}
\label{sec:approach_loop}

\textit{\textbf{Panel c}} of Figure~\ref{fig:approach} shows this step. Before repair begins, \approach{} writes the generated evidence to read-only files and mounts them in the agent container. The evidence consists of the broken-state error message, the first divergence when one is available, and the filtered library diff. The opening prompt presents the first divergence together with selected fragments of the filtered library diff. To choose these fragments, \approach{} scores each fragment of the filtered library diff by the symptom symbols that it contains, and a match in a library-test fragment counts twice as much as a match in a library-source fragment. The remaining evidence stays in the mounted files, and the agent decides whether to inspect it.

The agent then repairs the downstream project using the repository and the supplied evidence.

\subsection{\benchmark{}}
\label{sec:benchmark}

Evaluating a repair method for behavioral breaks requires more than a downstream project that fails after an upgrade. Because the upstream behavioral change may be intentional, tests that encode the old behavior can reject an otherwise correct adaptation. \benchmark{} is designed around this oracle problem, and every case provides a confirmed behavioral break, a reproducible three-state harness, and a reference fix whose test changes encode the intended post-upgrade behavior. We first explain why existing benchmarks do not meet these requirements (Section~\ref{sec:benchmark_motivation}), then describe how \benchmark{} is constructed and what it contains (Section~\ref{sec:benchmark_construction}), and finally explain how an agent works on \benchmark{} and how its repair is evaluated (Section~\ref{sec:benchmark_usage}).

\subsubsection{Why Existing Benchmarks Are Insufficient}
\label{sec:benchmark_motivation}

The repair task defined in Section~\ref{sec:bbc_definition} places three requirements on a benchmark. \textit{First}, a case must contain a confirmed behavioral break in a real downstream project. \textit{Second}, the case must provide a reference fix whose test changes encode the intended post-upgrade behavior, because an intentional behavioral change may invalidate tests that encode the old library behavior. \textit{Third}, the case must include a public harness that replays the intact, broken, and repaired states consistently. Table~\ref{tab:benchmark-scale} compares six existing dependency-upgrade benchmarks on these requirements. None of them focuses on behavioral breaking changes, and only TimeMachine-bench Verified has a confirmed behavioral subset, in which our audit finds 19 cases. In addition, BUMP and CompSuite provide no reference fixes, BDUpdater provides them for a minority of its cases, and DepBench and RepoRescue release no public harness. As of August 2026, no existing benchmark therefore meets all three requirements, so we construct \benchmark{}, whose size matches the three benchmarks that provide a reference fix for every case.

\begin{table}[t]
  \caption{Existing Dependency-Upgrade Benchmarks.}
  \label{tab:benchmark-scale}
  \footnotesize
  \begin{tabular}{lccccc}
    \toprule
    \textbf{Benchmark} & \textbf{Ecosystem} & \textbf{Cases} & \textbf{\begin{tabular}[b]{@{}c@{}}Behavioral\\ cases\end{tabular}} & \textbf{\begin{tabular}[b]{@{}c@{}}Reference\\ fix\end{tabular}} & \textbf{\begin{tabular}[b]{@{}c@{}}Public\\ harness\end{tabular}} \\
    \midrule
    \textbf{BUMP}~\cite{10589737} & Java & 571 & not labeled & no & yes \\
    \textbf{CompSuite}~\cite{10.1109/ASE56229.2023.00127} & Java & 123 & not labeled & no & yes \\
    \textbf{BDUpdater}~\cite{10.1145/3808116} & JavaScript & 84 & not labeled & 24 of 84 & yes \\
    \textbf{RepoRescue}~\cite{lin2026reporescueempiricalstudyllm} & Python & 146 & not labeled & yes & no \\
    \textbf{DepBench}~\cite{yang2026deprepairllmbasedsourcecoderepair} & \begin{tabular}[t]{@{}l@{}}Java, JavaScript,\\ Rust, Python\end{tabular} & 95 & excluded & yes & no \\
    \textbf{TimeMachine-bench Verified}~\cite{fujii-etal-2026-timemachine} & Python & 100 & 19 (our audit) & yes & yes \\
    \bottomrule
  \end{tabular}
\end{table}

\subsubsection{\benchmark{} Construction and Composition}
\label{sec:benchmark_construction}

\begin{figure}[t]
  \centering
  \includegraphics[width=\linewidth]{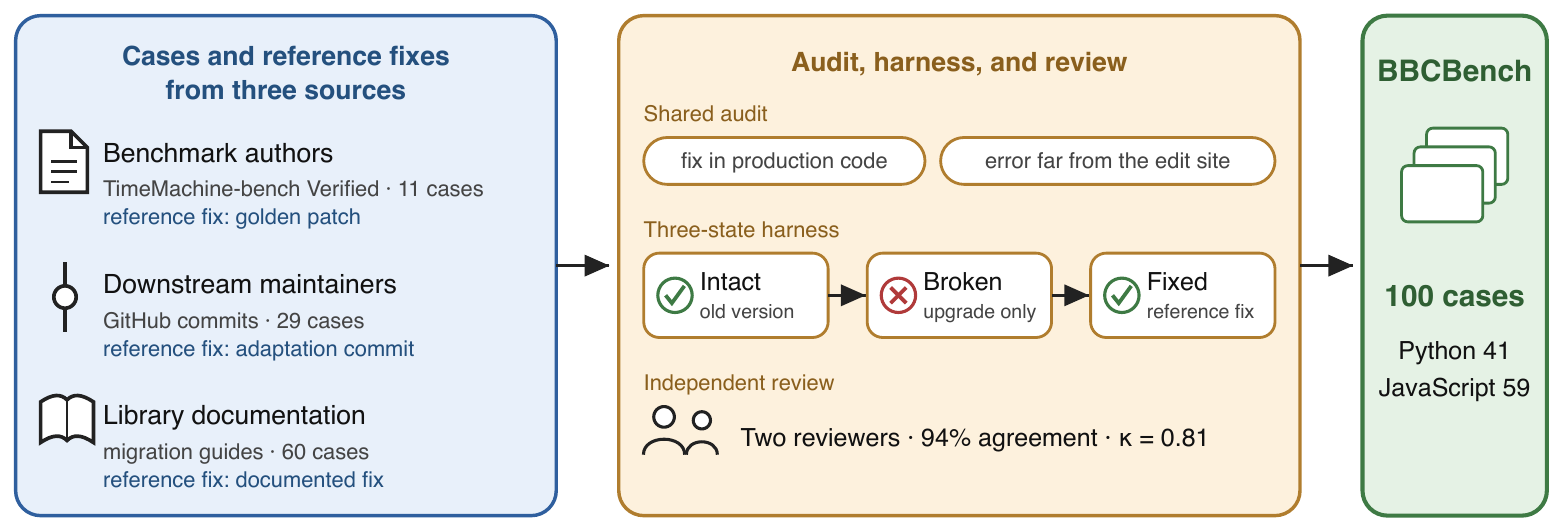}
  \caption{The construction pipeline of \benchmark{}.}
  \Description{A left-to-right pipeline with three columns. Column 1,
  cases and reference fixes from three sources, benchmark authors
  (TimeMachine-bench Verified, 11 cases, the golden patch as the
  reference fix), downstream maintainers (GitHub commits, 29 cases, the
  adaptation commit as the reference fix), and library documentation
  (migration guides, 60 cases, the documented fix as the reference fix).
  Column 2, audit, harness, and review, a shared audit with two
  criteria, a fix in production code and an error far from the edit
  site, a harness with three states, intact, broken, and fixed, and an
  independent review by two reviewers with 94\% agreement. Column 3,
  the result, \benchmark{} with 100 cases, 41 in Python and 59 in
  JavaScript.}
  \label{fig:construction}
\end{figure}

Figure~\ref{fig:construction} summarizes how \benchmark{} is built. We collect real-world cases and their reference fixes from three sources, and we audit, package, and review every case. We follow two steps below. 

\paragraph{Cases and Reference Fixes from Three Sources}
\label{sec:benchmark_sources}

Every candidate is a real downstream project on GitHub that a behavioral breaking change in a third-party dependency breaks. We collect candidates from three sources, and each source also provides the reference fix of its candidates in its own way. First, benchmark authors produce and verify fixes for their own cases. We read the golden patch of each of the 100 TimeMachine-bench Verified cases, and the golden patch serves as the reference fix of the case. Second, downstream maintainers write fixes when they adapt their projects to a new library version. We search GitHub for commits, pull requests, and issues of this kind, with queries such as \texttt{compatibility with numpy 2}, \texttt{adapt to pandas 2.0}, and \texttt{fix for <library> <version>}. We take the parent of such an adaptation commit as the project before the fix, and the edits of the commit to production code and tests form the reference fix. Third, library documentation records the standard fix for a behavioral change in the migration guide, changelog, or release notes. For each such change, we use GitHub code search and the dependents list of the npm registry to find downstream projects that call the changed API. We check out a commit at which the project still pins the old library version, upgrade only that library, and confirm that the test suite fails. We then apply the documented fix at the actual call sites as the reference fix.

None of these sources tells whether a candidate is behavioral, so we decide admission manually for every candidate from two observations. First, we read how the reference fix repairs the break. A fix that restores the old behavior or adapts the code to the new semantics indicates a behavioral breaking change, for example setting a default value explicitly. A fix that only renames an API, moves an import, or replaces a removed symbol indicates a syntactic breaking change, and we drop such candidates. Second, we run the tests after the upgrade. At least one test must fail on a wrong value or a runtime error that the changed behavior causes, not on a missing API. If the upgrade also removes or renames APIs, we first fix only those APIs and rerun the tests.

\paragraph{Audit, Harness, and Review}
\label{sec:benchmark_harness}

Every candidate then passes one shared audit with two criteria. First, the reference fix modifies production code, not only tests. Second, the error message must not point to the place that must be edited. We drop a case when the error trace points to a downstream line that the reference fix edits and one mechanical edit there repairs the break. Every admitted case is packaged as a reproducible harness that replays three states. In the \emph{intact} state, the project passes its tests with the old dependency version. In the \emph{broken} state, only the dependency is upgraded and at least part of the test suite fails. In the \emph{fixed} state, the reference fix is applied and the tests pass. The reference fix includes the test edits that encode the intended post-upgrade behavior. A container image fixes the runtime, and a prebuilt image or a saved lockfile fixes the package versions of each state. Automated gates then scan logs and code comments for text that leaks the fix, reject reference fixes that touch no production file, and flag test runs in which no test executes. Two reviewers who did not build the case also re-run the gates and the three states, and each independently judges whether the case is behavioral and meets the two criteria. The two reviewers initially agree on 94\% of the cases (Cohen's kappa 0.81) and resolve the remaining disagreements by discussion.

\paragraph{\benchmark{}}
\label{sec:benchmark_composition}

\benchmark{} consists of 100 cases in two ecosystems, 41 in Python and 59 in JavaScript. Each case packages the downstream project, the two library versions, the tests that fail after the upgrade, and the reference fix that adapts the downstream production code. By source, 11 cases come from the benchmark authors of TimeMachine-bench Verified, 29 from adaptation commits of downstream maintainers on GitHub, and 60 from library documentation, with the documented fix applied to real broken downstreams.

\subsubsection{How \benchmark{} Works}
\label{sec:benchmark_usage}

\benchmark{} provides a workspace for the agent to work in and an automated evaluation script that judges the result. The workspace is the broken state of a case: the downstream project before the fix, with the target dependency upgraded to \texttt{v\_new}, where at least one test fails on a wrong value or a runtime error that the changed behavior causes. In this workspace, the agent can read, run, and edit the project and its tests, and it can inspect the installed library. The evaluation script then decides whether the repair passes. At the end of a run, the script retains the agent's production-code changes but discards all changes to test files, preventing a run from passing by weakening or removing tests. The script then applies only the test changes from the reference fix. These tests encode the intended post-upgrade behavior, while the reference production-code changes remain hidden and unapplied, allowing the agent to produce an alternative valid repair. All tests execute in the broken state, with the dependency pinned to \texttt{v\_new}; a separate version check rejects any run that downgrades, removes, or replaces the target dependency. A run passes only if the version check succeeds, the project builds, and the complete post-upgrade test suite passes within the judging budget.

\subsection{Experimental Setup}
\label{sec:setup}

We evaluate \approach{} on \benchmark{} comparing Pure Agent, BDUpdater, and two ablation settings. The following paragraphs describe the agent, models, environment, and budgets; the baseline and ablation settings; the root-API labels; and the runs, metrics, and statistical tests.

\paragraph{Agent, Models, and Environment}
Every experiment uses the same agent, mini-swe-agent 2.4.6~\cite{minisweagent2025, yang2024sweagent}, a minimal agent framework whose only tool is the shell. The agent drives two open models, qwen3-coder~\cite{qwen3coder2025} and GLM-4.5-Air~\cite{glm45report}, through the OpenRouter API. Both models run at the default sampling temperature of 0.7 and keep the default values of all other sampling parameters. Each model call returns at most 4,096 tokens. All runs execute on one Linux virtual machine with 16 virtual CPUs and 62\,GB of memory, and every run works in its own Docker container that holds the downstream project with the upgraded library. The agent may not consult online resources about the downstream project itself, because the real fix of every case is public. Each shell command times out after 10 minutes. Every setting uses the same two models and the same budget: each run stops at 100 steps, 30 minutes, or US\$4, whichever the run reaches first, and cost is taken from the actual billing.

\paragraph{Baseline Settings}
We compare \approach{} with two settings. (1) The \textbf{Pure Agent} setting, which shows what the agent repairs on its own, without added instructions. In this setting, the agent receives only the broken downstream repository and its error message. The agent can inspect the repository and the installed library but receives no upgrade evidence. (2) The \textbf{BDUpdater }setting, which replicates the method design of the original tool~\cite{10.1145/3808116}, adapted to \benchmark{} by mounting the released sources of both library versions and the upstream repository instead of network lookups and by dropping its dependency-bump step. In this setting, the agent itself mines changelogs, release notes, and commit messages into a list of breaking changes, verifies each entry against the source, and adapts the affected call sites. 

\paragraph{Ablation Settings}
Two further settings serve the ablation study and keep the repair procedure of \approach{} (Section~\ref{sec:approach_loop}). Each ablation setting keeps one of the two evidence items and removes the other. \approach{} (without first divergence) filters the library diff with the keys that \approach{} uses when no divergence is available (Section~\ref{sec:approach_filter}) and gives the agent that filtered library diff. \approach{} (without library diff) gives the agent the first divergence but no library diff. The agent therefore learns which API changed and how its return value changed, but not what the library changed inside that API.

\paragraph{Root-API Labels}
Every case carries a labeled root API, the upstream API whose behavioral change causes the break. We label it from the reference fix and the upstream change of each case. An LLM assigns the label under a written protocol, two authors then independently label every case, and they agree on 92.0\% of the cases (Cohen's kappa 0.83). This LLM is Claude Haiku 4.5~\cite{claudehaiku45}, and the same model proposes every other LLM-assigned label in the paper. Section~\ref{sec:results} uses this label to judge whether a run identifies the root API and whether the evidence points to it.

\paragraph{Runs, Metrics, and Statistical Tests}
Each setting repeats the full benchmark three times, because the outcome of a single run varies between repetitions of the same case. A round is one repetition over all 100 cases, and a run is one case in one round, so each setting produces 300 runs per model. The primary metric is the average pass rate over the three rounds, and the cases passing in all three runs, in the best round, in at least two runs, and in any run (pass@3) give robustness views. We test each difference between two settings at the case level and treat the three runs of a case as one unit. A paired bootstrap over the 100 cases gives the 95\% confidence interval of the pass-rate difference in percentage points, a Wilcoxon signed-rank test compares the per-case pass counts, and an exact McNemar test compares the cases that only one setting repairs in all three runs. The result tables report these two McNemar counts as the A only and B only columns. Within each model, the $p$-values of each result table are Holm-corrected over its two comparisons of \approach{}.

\section{Results}
\label{sec:results}

We answer four research questions to evaluate our approach:
\begin{itemize}
  \item \textbf{RQ1 (Pure Agent Behavior and Failure Patterns):} What upstream information do LLM-based agents consult when they repair a behavioral breaking change under the Pure Agent setting, and what failure patterns do their trajectories present?
  \item \textbf{RQ2 (Effectiveness of \approach{}):} To what extent does \approach{} improve the repair performance of agents on behavioral breaking changes compared with two baselines (i.e.,  Pure Agent and BDUpdater), and at what cost in agent effort?
  \item \textbf{RQ3 (Effect of Different Evidence Configurations):} How does the repair performance of \approach{} change when the library diff is filtered without the first divergence, or when the first divergence is supplied without the library diff?
  \item \textbf{RQ4 (Root-API Identification and Repair Outcomes):} How often does each setting identify the root API (RQ4.1), does identifying it predict a successful repair (RQ4.2), and does the gain of \approach{} concentrate on the cases whose evidence points to the labeled root API (RQ4.3)?
\end{itemize}

\subsection{RQ1: Pure Agent Behavior and Failure Patterns}
\label{sec:rq1}

RQ1 examines how the agent behaves under the Pure Agent setting (Section~\ref{sec:setup}), in which it repairs a behavioral break without upgrade-specific evidence.

\paragraph{Use of Upgrade-Specific Information}

We classify the shell commands of all 300 trajectories per model with a fixed rule, which counts only the commands that read materials of the upgraded library, and record two types of consulted information: \emph{textual information} (README, changelog, and migration guides) and \emph{code-level information} (the installed upstream source and its tests). On a random sample of 100 trajectories, the rule agrees with the manual labels on an average of 87.0\% of them (Cohen's kappa 0.83). Table~\ref{tab:info-consultation} reports the distribution.

Under both models, the most common behavior is to consult neither type of information, at 61.3\% of qwen3-coder runs and 83.0\% of GLM-4.5-Air runs; these runs rely only on the error message and trial-and-error edits. The runs that do consult something read code more often than text. Even these runs stop short of comparing the two library versions: no qwen3-coder run does so through source code or tests. Pure Agent runs therefore rarely obtain the upgrade-specific evidence that connects the downstream failure to the upstream change.

\paragraph{Failure Patterns}

We classify every failing run of the Pure Agent setting by where the repair breaks down. The classification asks whether the trajectory shows that the agent identifies the labeled root API of the case (Section~\ref{sec:setup}) and understands how its behavior changes in the upgrade. An LLM judge assigns the category from the commands of the trajectory under a written protocol, with the contents of every edit hidden. Two authors then label every failing run independently, agree on 92.6\% of the runs (Cohen's kappa 0.85), and resolve the disagreements by discussion. The three categories are:
\begin{itemize}
  \item \textbf{Root API not identified.} The run never identifies the root API, and instead patches at the error site, tries unrelated options, or attributes the failure to the downstream code.
  \item \textbf{Incorrect adaptation.} The run identifies the root API but writes a wrong fix.
  \item \textbf{Execution failure.} The run identifies the root API but introduces syntax errors or violates a task constraint of Section~\ref{sec:bbc_definition}.
\end{itemize}

\begin{table}[t]
  \begin{minipage}[t]{0.48\textwidth}
    \centering
    \caption{Distribution of Information Consultation Behavior over the Pure Agent Runs.}
    \label{tab:info-consultation}
    \footnotesize
    \setlength{\tabcolsep}{3pt}
    \begin{tabular}{@{}lcc@{}}
      \toprule
      \textbf{Consulted information} & \textbf{qwen3-coder} & \textbf{GLM-4.5-Air} \\
      \midrule
      \textbf{Both types}      & 7.3  & 6.3  \\
      \textbf{Code-level only} & 26.0 & 7.7  \\
      \textbf{Textual only}    & 5.3  & 3.0  \\
      \textbf{Neither}         & 61.3 & 83.0 \\
      \bottomrule
    \end{tabular}
  \end{minipage}\hfill
  \begin{minipage}[t]{0.48\textwidth}
    \centering
    \caption{Distribution of Failure Patterns over the Failing Pure Agent Runs.}
    \label{tab:failure-categories}
    \footnotesize
    \setlength{\tabcolsep}{3pt}
    \def\rowstrut{\rule[-\dimexpr\dp\strutbox+\normalbaselineskip/6\relax]{0pt}{\dimexpr\ht\strutbox+\dp\strutbox+\normalbaselineskip/3\relax}}
    \begin{tabular}{@{}lcc@{}}
      \toprule
      \textbf{Failure pattern} & \textbf{qwen3-coder} & \textbf{GLM-4.5-Air} \\
      \midrule
      \rowstrut\textbf{Root API not identified} & 66.9 & 69.2 \\
      \rowstrut\textbf{Incorrect adaptation}    & 27.8 & 29.2 \\
      \rowstrut\textbf{Execution failure}       & 5.3  & 1.7  \\
      \bottomrule
    \end{tabular}
  \end{minipage}
\end{table}

Table~\ref{tab:failure-categories} reports the distribution over the 151 failing Pure Agent runs of qwen3-coder and the 120 of GLM-4.5-Air. Not identifying the root API is the dominant failure category under both models, at 66.9\% and 69.2\%, so the repair of Pure Agent breaks down mainly at the diagnosis rather than at the edit. Incorrect adaptation accounts for most of the remaining runs, and execution failures are rare. The model that consults less, GLM-4.5-Air, also has a larger share of failures in diagnosis.

\begin{tcolorbox}[colback=black!5!white, colframe=black!70!white, colbacktitle=black!15!white, coltitle=black, title=\textbf{Summary of RQ1}]
\setlength{\leftmargini}{1em}\hyphenpenalty=10000
\begin{itemize}
  \item \textbf{Repository access does not ensure evidence retrieval.} Most Pure Agent runs consult neither textual nor code-level information about the upgraded library.
  \item \textbf{Diagnosis is the dominant observed failure pattern.} Failure to identify the labeled root API accounts for the majority of failing runs under both models.
\end{itemize}
\end{tcolorbox}

\subsection{RQ2: Effectiveness of \approach{}}
\label{sec:rq2}

To answer this RQ, we compare the repair performance of \approach{} with that of Pure Agent and BDUpdater~\cite{10.1145/3808116}, test the differences for significance, and compare the agent effort that each setting spends per successful repair.

\begin{table}[!t]
  \caption{Repair Outcomes over Three Rounds.}
  \label{tab:rq2-pass-rate}
  \footnotesize
  \setlength{\tabcolsep}{2pt}
  \begin{tabular}{@{}lcccccccc@{}}
    \toprule
    & \multicolumn{4}{c}{\textbf{qwen3-coder}} & \multicolumn{4}{c}{\textbf{GLM-4.5-Air}} \\
    \cmidrule(lr){2-5} \cmidrule(lr){6-9}
    \textbf{Metric} & \textbf{\approach{}} & \textbf{BDUpdater} &
    \textbf{\shortstack{Pure\\Agent}} & \textbf{Gain} &
    \textbf{\approach{}} & \textbf{BDUpdater} &
    \textbf{\shortstack{Pure\\Agent}} & \textbf{Gain} \\
    \midrule
    \textbf{Average pass rate}                 & \textbf{58.0} & 38.0 & 49.7 & \textbf{+8.3 ($\uparrow$17\%)} & \textbf{67.3} & 61.7 & 60.0 & \textbf{+7.3 ($\uparrow$12\%)} \\
    \textbf{Pass in all 3 runs}       & \textbf{42}   & 15   & 33   & \textbf{+9 ($\uparrow$27\%)} & \textbf{56} & 47 & 42 & \textbf{+14 ($\uparrow$33\%)} \\
    \textbf{Pass in the best round} & \textbf{63}   & 46   & 52   & \textbf{+11 ($\uparrow$21\%)} & \textbf{71} & 64 & 64 & \textbf{+7 ($\uparrow$11\%)} \\
    \textbf{Pass in $\geq$2 runs}     & \textbf{58}   & 35   & 48   & \textbf{+10 ($\uparrow$21\%)} & \textbf{70} & 60 & 63 & \textbf{+7 ($\uparrow$11\%)} \\
    \textbf{Pass in any run (pass@3)} & \textbf{74}   & 64   & 68   & \textbf{+6 ($\uparrow$9\%)} & 76 & \textbf{78} & 75 & +1 ($\uparrow$1\%) \\
    \bottomrule
  \end{tabular}
\end{table}

\begin{table}[!t]
  \caption{Statistical Tests of the Pass-Rate Gains of \approach{}.}
  \label{tab:rq2-stats}
  \footnotesize
  \begin{tabular}{llcccc}
    \toprule
    \textbf{Model} & \textbf{A vs.\ B} & \textbf{95\% CI} & \textbf{Wilcoxon $p$} & \textbf{A only / B only} & \textbf{McNemar $p$} \\
    \midrule
    \textbf{qwen3-coder} & \textbf{\approach{} vs.\ Pure Agent}    & [+1.7, +15.3]  & 0.020   & 13 / 4 & 0.049 \\
                         & \textbf{\approach{} vs.\ BDUpdater} & [+12.7, +27.3] & $<$0.001 & 31 / 4  & $<$0.001 \\
    \midrule
    \textbf{GLM-4.5-Air} & \textbf{\approach{} vs.\ Pure Agent}    & [+1.3, +13.7] & 0.072 & 19 / 5  & 0.014 \\
                         & \textbf{\approach{} vs.\ BDUpdater} & [$-$2.3, +13.7] & 0.130 & 22 / 13 & 0.175 \\
    \bottomrule
  \end{tabular}
\end{table}

\begin{table}[!t]
\centering
\caption{Budget Use over 300 Runs per Model.}
\label{tab:budget}
  \footnotesize
\begin{tabular}{llrrrrr}
\toprule
\textbf{Model} & & \textbf{\approach{}} & \textbf{\shortstack{vs.\\BDUpdater}}
& \textbf{\shortstack{vs. Pure\\Agent}} & \textbf{BDUpdater}
& \textbf{\shortstack{Pure\\Agent}} \\
\midrule
\textbf{qwen3-coder} & Agent steps / successful repair
& 91.9 & \textbf{$\downarrow$58\%} & \textbf{$\downarrow$27\%} & 219.7 & 125.3 \\
& Cost / successful repair (cents)
& 9.6 & \textbf{$\downarrow$58\%} & \textbf{$\downarrow$2\%} & 22.7 & 9.8 \\
& Budget-exhausted runs (\%)
& 27.9 & \textbf{$\downarrow$51\%} & \textbf{$\downarrow$19\%} & 57.3 & 34.3 \\
\midrule
\textbf{GLM-4.5-Air} & Agent steps / successful repair
& 84.0 & \textbf{$\downarrow$29\%} & \textbf{$\downarrow$15\%} & 117.6 & 98.8 \\
& Cost / successful repair (cents)
& 5.4 & \textbf{$\downarrow$27\%} & \textbf{$\downarrow$7\%} & 7.4 & 5.8 \\
& Budget-exhausted runs (\%)
& 26.6 & \textbf{$\downarrow$28\%} & \textbf{$\downarrow$21\%} & 36.7 & 33.7 \\
\bottomrule
\end{tabular}
\end{table}

Table~\ref{tab:rq2-pass-rate} compares the repair outcomes of the
three settings. \approach{} outperforms
Pure Agent under every view with both models, raising the average pass rate by 17\% with qwen3-coder and
by 12\% with GLM-4.5-Air. The size
of the gain follows the strictness of the view. Under the
strictest view, the cases that pass in all three runs, the gain
grows to 27\% and 33\%. These results show that \approach{} improves
repair robustness at scale: its advantage lies not primarily in reaching
cases that Pure Agent never repairs, but in repairing the same cases
consistently across repeated runs. Under both models, \approach{} also
passes more cases than both Pure Agent and BDUpdater in each of the three rounds, further demonstrating the consistency
of its improvement rather than an average driven by a single strong round.

BDUpdater never exceeds \approach{} in average pass rate. With qwen3-coder it stays below even Pure Agent under every view, and with GLM-4.5-Air it stays close to Pure Agent and passes \approach{} only under pass@3. BDUpdater repairs many cases in at least one run but few of them in all three, which makes it the least stable of the three settings. A reading of the 54 failing runs of this setting in the first round under qwen3-coder shows two causes, and neither is missing documentation. About half of the runs spend most of the shared budget on mining and end with no repair action or with edits only in the final steps. About a third of the runs see the text that documents the root API yet edit the wrong file or apply a wrong fix, because documentation records what changes in the library, not how the downstream code must change.

Table~\ref{tab:rq2-stats} tests the gains of
Table~\ref{tab:rq2-pass-rate}. In the McNemar test, \approach{} is
significantly better than Pure Agent under both models and better
than BDUpdater with qwen3-coder. The confidence interval
of the pass-rate difference over Pure Agent excludes zero under
both models, and with qwen3-coder the Wilcoxon test agrees for both
comparisons.

Table~\ref{tab:budget} reports how the three settings use the same
per-run budget.
We compute agent steps and cost per successful repair as the total
over all 300 runs divided by the number of passing runs.
A run counts as budget-exhausted when it reaches the step cap.
\approach{} improves the pass rate while using less of the
budget. It needs 15\% to 27\% fewer agent steps per successful
repair than Pure Agent and exhausts the step budget less often
under both models. Against BDUpdater, \approach{} needs
29\% to 58\% fewer agent steps and 27\% to 58\% less cost per
successful repair. BDUpdater also raises the
cost per successful repair above Pure Agent, by 132\% with
qwen3-coder and by 28\% with GLM-4.5-Air. Mining and repair share
one budget, and BDUpdater reaches the step cap in 57.3\% of its runs
under qwen3-coder, so documentation search consumes the
budget that the repair needs.

The case hubot-coffeepoll in Figure~\ref{fig:bbc-consequence} shows how the evidence obtained by \approach{} changes a run. During the repair process of \approach{}, the first divergence shows that \texttt{\_.sample()} returns three elements under the old version and one object under the new version. The agent names it as the cause within five steps and applies the one-line reference fix after 26 steps. Pure Agent instead keeps editing \texttt{pickWinner}, the function that the failing assertions point to, and reaches the step cap without changing the \texttt{\_.sample()} call. BDUpdater finds only a link to the online changelog, never reads the \texttt{\_.sample()} call, and also reaches the cap.

\begin{tcolorbox}[colback=black!5!white, colframe=black!70!white, colbacktitle=black!15!white, coltitle=black, title=\textbf{Summary of RQ2}]
\setlength{\leftmargini}{1em}\hyphenpenalty=10000
\begin{itemize}
  \item \textbf{\approach{} outperforms both Pure Agent and BDUpdater.} It also passes more cases than both in every round and the gain is significant.
  \item \textbf{\approach{} can reduce the Agent cost and budget-exhausted runs per successful repair.}
  \item \textbf{BDUpdater brings no consistent gain over Pure Agent and consumes more agent cost.} 
\end{itemize}
\end{tcolorbox}

\subsection{RQ3: Effect of Different Evidence Configurations}
\label{sec:rq3}

To answer this RQ, we compare \approach{} with Pure Agent and with the two ablation settings defined in Section~\ref{sec:setup}, \approach{} (without first divergence) and \approach{} (without library diff).

\begin{table}[!t]
  \caption{Ablation of \approach{} over Three Rounds under Both Models.}
  \label{tab:rq3-ablation}
  \setlength{\tabcolsep}{2pt}
  \footnotesize
  \begin{tabular}{@{}lcccccccc@{}}
    \toprule
    & \multicolumn{4}{c}{\textbf{qwen3-coder}} & \multicolumn{4}{c}{\textbf{GLM-4.5-Air}} \\
    \cmidrule(lr){2-5} \cmidrule(lr){6-9}
    \textbf{Metric} & \textbf{\approach{}} & \textbf{\shortstack{\approach{}\\(without\\first\\divergence)}} &
    \textbf{\shortstack{\approach{}\\(without\\library\\diff)}} & \textbf{\shortstack{Pure\\Agent}} &
    \textbf{\approach{}} & \textbf{\shortstack{\approach{}\\(without\\first\\divergence)}} &
    \textbf{\shortstack{\approach{}\\(without\\library\\diff)}} & \textbf{\shortstack{Pure\\Agent}} \\
    \midrule
    \textbf{Average pass rate}                 & \textbf{58.0} & 53.7 & 51.3 & 49.7 & \textbf{67.3} & 61.0 & 62.7 & 60.0 \\
    \textbf{Pass in all 3 runs}       & \textbf{42}   & 34   & 38   & 33   & \textbf{56}   & 48   & 47   & 42   \\
    \textbf{Pass in the best round} & \textbf{63}   & 56   & 53   & 52   & \textbf{71}   & 62   & 66   & 64   \\
    \textbf{Pass in $\geq$2 runs}     & \textbf{58}   & 56   & 52   & 48   & \textbf{70}   & 61   & 65   & 63   \\
    \textbf{Pass in any run (pass@3)} & \textbf{74}   & 71   & 64   & 68   & \textbf{76}   & 74   & \textbf{76}   & 75   \\
    \bottomrule
  \end{tabular}
\end{table}

\begin{table}[!t]
  \caption{Statistical Tests of \approach{} against Each Ablation Setting.}
  \label{tab:rq3-stats}
  \footnotesize
  \begin{tabularx}{\linewidth}{lXcccc}
    \toprule
    \textbf{Model} & \textbf{A vs.\ B} & \textbf{95\% CI} & \textbf{Wilcoxon $p$} & \textbf{A only / B only} & \textbf{McNemar $p$} \\
    \midrule
    \textbf{qwen3-coder} & \textbf{\approach{} vs.\ \approach{}\newline (without first divergence)} & [$-$1.7, +10.3] & 0.189 & 16 / 8  & 0.304 \\
                         & \textbf{\approach{} vs.\ \approach{}\newline (without library diff)} & [+1.3, +12.3]   & 0.044 & 9 / 5   & 0.424 \\
    \midrule
    \textbf{GLM-4.5-Air} & \textbf{\approach{} vs.\ \approach{}\newline (without first divergence)} & [+0.0, +12.7]   & 0.090 & 18 / 10 & 0.185 \\
                         & \textbf{\approach{} vs.\ \approach{}\newline (without library diff)} & [+0.0, +9.7]    & 0.090 & 14 / 5  & 0.128 \\
    \bottomrule
  \end{tabularx}
\end{table}
Table~\ref{tab:rq3-ablation} shows that \approach{} achieves a higher average pass rate than both ablation settings under each model. It also matches or exceeds both ablation settings across all five views. In contrast, neither ablation setting consistently improves on Pure Agent, and each falls below it under at least one view. These results suggest that both components contribute to \approach{}'s effectiveness.

Table~\ref{tab:rq3-stats} tests the gain of \approach{} over each ablation setting. The gain over \approach{} (without library diff) is significant under qwen3-coder ($p=0.044$), and the other three comparisons point in the same direction.

\begin{tcolorbox}[colback=black!5!white, colframe=black!70!white, colbacktitle=black!15!white, coltitle=black, title=\textbf{Summary of RQ3}]
\hyphenpenalty=10000
\textbf{The gain comes from how the evidence is built.} A library diff filtered without the first divergence, and the first divergence without any library diff, both stay close to Pure Agent. Only \approach{}, where the first divergence keys the filtered library diff and both reach the agent, lifts both models above Pure Agent under every view.
\end{tcolorbox}

\subsection{RQ4: Root-API Identification and Repair Outcomes}
\label{sec:rq4}

To answer RQ4, we examine when the evidence generated by \approach{} is most useful for repair. RQ4.1 measures how often each setting enables the agent to identify the relevant upstream API. RQ4.2 examines the association between identifying that API and successful repair. RQ4.3 compares the improvement of \approach{} on the cases whose generated evidence points to the labeled root API and on all cases.

\subsubsection{RQ4.1: Agent Identification of the Root API}
\label{sec:rq4_locating}

Section~\ref{sec:rq1} labels whether a run identifies the root API only for the failing runs of Pure Agent. We extend this identification label to every run of the three settings of RQ2 under the same protocol, and two authors again review every run independently (95.5\% agreement, Cohen's kappa 0.85). Table~\ref{tab:rq4-locating} reports the results. Pure Agent identifies the root API in about half of its runs under both models. Both evidence-based settings raise that rate by 20\% to 22\% relative to Pure Agent. \approach{} raises it by 22\% under both models, and BDUpdater raises it slightly less under both models. Both forms of upstream evidence therefore help the agent identify the upstream behavior.

\begin{table}[t]
  \caption{Rate at Which Agents Identify the Labeled Root API.}
  \label{tab:rq4-locating}
  \footnotesize
  \begin{tabular}{lcccc}
    \toprule
    & \multicolumn{2}{c}{\textbf{qwen3-coder}} & \multicolumn{2}{c}{\textbf{GLM-4.5-Air}} \\
    \cmidrule(lr){2-3} \cmidrule(lr){4-5}
    \textbf{Setting} & \textbf{Identified} & \textbf{Gain} & \textbf{Identified} & \textbf{Gain} \\
    \midrule
    \textbf{Pure Agent}     & 50.0 & --    & 51.0 & --    \\
    \textbf{BDUpdater} & 60.3 & +10.3 ($\uparrow$21\%) & 61.0 & +10.0 ($\uparrow$20\%) \\
    \textbf{\approach{}}  & 61.0 & +11.0 ($\uparrow$22\%) & 62.0 & +11.0 ($\uparrow$22\%) \\
    \bottomrule
  \end{tabular}
\end{table}

\subsubsection{RQ4.2: Association between Root-API Identification and Repair}
\label{sec:rq4_locating_repair}

\begin{table}[!t]
  \caption{Identifying the Root API and Repair Outcome under Both Models.}
  \label{tab:rq4-locate-pass}
  \footnotesize
  \begin{tabular}{lcccccc}
    \toprule
    & \multicolumn{3}{c}{\textbf{qwen3-coder}} & \multicolumn{3}{c}{\textbf{GLM-4.5-Air}} \\
    \cmidrule(lr){2-4} \cmidrule(lr){5-7}
    & \textbf{Pure Agent} & \textbf{BDUpdater} & \textbf{\approach{}} & \textbf{Pure Agent} & \textbf{BDUpdater} & \textbf{\approach{}} \\
    \midrule
    \multicolumn{7}{l}{\emph{Pass rate by whether the run identifies the root API}} \\
    \quad \textbf{Identified}       & 66.7 ($\uparrow$104\%) & 53.0 ($\uparrow$251\%) & 70.5 ($\uparrow$83\%) & 75.8 ($\uparrow$74\%) & 77.6 ($\uparrow$111\%) & 77.4 ($\uparrow$52\%) \\
    \quad \textbf{Not identified}   & 32.7\phantom{ ($\uparrow$104\%)} & 15.1\phantom{ ($\uparrow$251\%)} & 38.5\phantom{ ($\uparrow$83\%)} & 43.5\phantom{ ($\uparrow$74\%)} & 36.8\phantom{ ($\uparrow$111\%)} & 50.9\phantom{ ($\uparrow$52\%)} \\
    \midrule
    \multicolumn{7}{l}{\emph{Rate of identifying the root API by outcome}} \\
    \quad \textbf{Passing runs} & 67.1\phantom{ ($\uparrow$104\%)} & 84.2\phantom{ ($\uparrow$251\%)} & 74.1\phantom{ ($\uparrow$83\%)} & 64.4\phantom{ ($\uparrow$74\%)} & 76.8\phantom{ ($\uparrow$111\%)} & 71.3\phantom{ ($\uparrow$52\%)} \\
    \quad \textbf{Failing runs} & 33.1\phantom{ ($\uparrow$104\%)} & 45.7\phantom{ ($\uparrow$251\%)} & 42.9\phantom{ ($\uparrow$83\%)} & 30.8\phantom{ ($\uparrow$74\%)} & 35.7\phantom{ ($\uparrow$111\%)} & 42.9\phantom{ ($\uparrow$52\%)} \\
    \bottomrule
  \end{tabular}

  \vspace{\baselineskip}

  \caption{Gain of \approach{} over Pure Agent on the Cases Whose
  Evidence Points to the Root API, under Both Models.}
  \label{tab:rq4-gain}
  \footnotesize
  \setlength{\tabcolsep}{3pt}
  \begin{tabular}{lccccccc}
    \toprule
    & & \multicolumn{3}{c}{\textbf{qwen3-coder}} & \multicolumn{3}{c}{\textbf{GLM-4.5-Air}} \\
    \cmidrule(lr){3-5} \cmidrule(lr){6-8}
    \textbf{Case set} & $\mathbf{n}$ & \textbf{\shortstack{Pure\\Agent}} & \textbf{\approach{}} & \textbf{Gain} &
    \textbf{\shortstack{Pure\\Agent}} & \textbf{\approach{}} & \textbf{Gain} \\
    \midrule
    \textbf{Evidence points to the root API} & 51  & 53.6 & 64.1 & +10.5 ($\uparrow$20\%) & 61.4 & 71.2 & +9.8 ($\uparrow$16\%) \\
    \textbf{All cases}                       & 100 & 49.7 & 58.0 & +8.3 ($\uparrow$17\%) & 60.0 & 67.3 & +7.3 ($\uparrow$12\%) \\
    \bottomrule
  \end{tabular}

  \vspace{\baselineskip}

  \caption{Reasons Why the Evidence Does Not Point to the Labeled Root API.}
  \label{tab:rq4-miss}
  \footnotesize
  \setlength{\tabcolsep}{6pt}
  \begin{tabularx}{\textwidth}{lXc}
    \toprule
    \textbf{Reason} & \textbf{Definition} & \textbf{\%} \\
    \midrule
    \textbf{Call not recorded} & The two runs do not both record the call to the root API, e.g., because the test runner loads the library through its own module loader or a wrapper hides the call. & 55.1 \\
    \textbf{Divergence ranked lower} & The call to the root API diverges, but another divergence ranks higher. & 22.4 \\
    \textbf{Change not in return values} & The behavior change alters no recorded return value, e.g., it acts through an operator, a setting, or a callback. & 22.4 \\
    \bottomrule
  \end{tabularx}
\end{table}

Table~\ref{tab:rq4-locate-pass} splits the runs of each setting by
whether the run identifies the root API and whether it passes. The upper half reports the pass rate by whether the run identifies the root API, with the relative increase of the identified runs over the not identified runs in parentheses. The lower half reports the rate of identifying the root API by outcome, separately for the passing runs and the failing runs.

The runs that identify the root API pass far more often in all six groups, by 52\% to 251\% relative to the runs that do not identify it. Passing runs also identify it more often than failing runs, in Pure Agent as well. Identifying the root API is therefore associated with a higher pass rate in every setting, not only where \approach{} supplies the evidence. \approach{} also has the highest pass rate among the runs that do not identify the root API under both models. The evidence of \approach{} carries the old and the new behavior, so an agent that never completes the diagnosis can still repair from it.

\subsubsection{RQ4.3: Gain on Cases Whose Evidence Points to the Root API}
\label{sec:rq4_gain_concentration}

We next examine whether the gain of \approach{} concentrates on the cases whose evidence points to the labeled root API. The \textbf{evidence points to the labeled root API} when either the first divergence or the filtered library diff identifies that API. We label each case for this property (Section~\ref{sec:setup}) from the calls recorded in the two runs. An LLM assigns the label under a written protocol, two authors then independently label every case, and they agree on 95.0\% of the cases (Cohen's kappa 0.89). The evidence points to the labeled root API in 51 of the 100 benchmark cases.

Table~\ref{tab:rq4-gain} compares the average pass rate of \approach{} and Pure Agent on those 51 cases and on all 100. Under both models, \approach{} passes more often on the 51 cases, and its gain over Pure Agent is larger there: 20\% instead of 17\% under qwen3-coder and 16\% instead of 12\% under GLM-4.5-Air. Section~\ref{sec:discussion} analyzes why the evidence misses the labeled root API in the remaining 49 cases.

\begin{tcolorbox}[colback=black!5!white, colframe=black!70!white, colbacktitle=black!15!white, coltitle=black, title=\textbf{Summary of RQ4}]
\setlength{\leftmargini}{1em}\hyphenpenalty=10000
\begin{itemize}
  \item \textbf{\approach{} raises the rate of identifying the root API, and identification is associated with repair success.} \approach{} identifies the labeled root API more often than Pure Agent under both models, and identified runs pass more often under every setting.
  \item \textbf{\approach{} gains most on the cases whose evidence points to the labeled root API.} Under both models, the gain over Pure Agent is larger on these cases than on the full benchmark.
\end{itemize}
\end{tcolorbox}

\section{Discussion}
\label{sec:discussion}

The results raise two implications beyond the RQs. The first is why the evidence sometimes misses the root API, and the second is what the findings imply for supplying evidence to agents.

\paragraph{Why the Evidence Sometimes Misses the Root API}
In RQ4.3, the evidence of \approach{} points to the labeled root API in 51 of the 100 cases. To find out why the evidence misses the root API in the other 49 cases, we label each of these cases with a reason, based on the calls recorded in the two runs. An LLM assigns the reason under a written protocol, two authors then independently label every case, and they agree on 91.8\% of the cases (Cohen's kappa 0.86). Table~\ref{tab:rq4-miss} lists the three reasons. In more than half of the 49 cases, the two runs do not both record the call to the root API. The other cases split evenly between two reasons: another divergence ranks higher than the call to the root API, or the behavior change alters no recorded return value.

The three reasons differ in whether differential execution can observe the break. \approach{} compares the return values of calls that enter the library, so it can observe a behavioral change that alters the result of an API call. In the cases with the first two reasons, the change still alters the result of a call, but the recorder does not record that call or the ranking rule does not rank it first. Therefore, two directions can help the evidence point to the root API in more cases: capturing calls that pass through the module loaders of test runners or through wrappers, and improving the ranking rule. In the cases with the third reason, a behavioral change modifies a general library rule, such as the type promotion rule of numpy 2, and takes effect through operators, settings, or callbacks that the recorder does not observe. Extending the recorder to operator results and attribute reads is a direction for covering part of these cases.

\paragraph{Showing the Behavior Change Helps beyond Identifying the Root API}
Among the runs that never identify the root API, \approach{} still has the highest pass rate under both models (RQ4.2). The gain of \approach{} therefore does not come only from identifying the root API. The first divergence also shows how the behavior of the call changes, because it records the return value of the call under the old and the new library version. In the running example of Figure~\ref{fig:approach}, the first divergence shows that \texttt{format\_date} returns \texttt{'11/08/2026'} instead of \texttt{'2026-08-11'} after the upgrade. This behavior change suggests the adaptation that the downstream code needs, even in a run that never identifies the root API. Evidence for repair agents should therefore show the old and the new behavior of the root API, not only its name. A tool that prepares such evidence also fits into continuous integration, because both inputs of differential execution, the project before and after the upgrade and the failing test, exist in every dependency-upgrade pull request.

\section{Threats to Validity}
\label{sec:threats}

\paragraph{Construct Validity}
The pass rate is an upper bound on repair correctness, because the tests of the reference fix do not cover every behavior. All settings share the same evaluation script, so the differences between settings are less affected than the absolute rates. The failure categories of RQ1 and the identification labels of RQ4.1 come from an LLM judge that follows a written protocol, and two authors review every label independently. The root-API labels are heuristic candidates rather than guaranteed root causes, so RQ4 measures whether the evidence points to the labeled root API and how that relates to the repair outcome, not the accuracy of root-cause localization.

\paragraph{Internal Validity}
The outcome of a single run varies between repetitions of the same setting, so every number in the paper is a three-round average. With 100 cases the statistical tests have limited power, the McNemar test for \approach{} against Pure Agent rests on 17 to 24 discordant cases, and the ablation has less power still. The evidence of \approach{} is generated once before the repair starts, so the agent cannot rerun differential execution or re-filter the library diff during the repair. Because \approach{} packages its evidence with an evidence-consumption procedure, the comparison with Pure Agent estimates the effect of the complete workflow rather than the isolated effect of the two evidence artifacts. A controlled comparison under an identical prompt remains necessary to attribute the gain solely to evidence selection.

\paragraph{External Validity}
\benchmark{} contains 41 Python and 59 JavaScript cases, and the evaluation uses one agent framework and two open models, so results by library, language, or model rest on small samples. We choose open, low-priced models because realistic deployment repairs many projects at every upgrade (Section~\ref{sec:introduction}). The gains of \approach{} are therefore not measured on higher-priced proprietary models. Differential execution records only the calls that enter the library from outside it, so when it reports no divergence, the filtered library diff carries the diagnosis alone. That filtered library diff states the new expected value only when the library revises its own tests, which Mostafa et al.\ find for 90.2\% of the behavioral backward incompatibilities they detect~\cite{10.1145/3092703.3092721}. This share is an upper bound for our setting, because those incompatibilities are found by running the library tests themselves, which cover them by construction. Some reference fixes come from public repositories and may have appeared in model training data, although agents cannot access online resources during repair and never see the reference fix. Future work should evaluate on temporally held-out upgrades and newly collected cases.

\section{Related Work}
\label{sec:related}

This work builds on two areas of prior research, empirical studies and detection tools for behavioral breaking changes (Section~\ref{sec:related_bbc}), and methods that repair breaking changes (Section~\ref{sec:related_repair}).

\subsection{Behavioral Breaking Changes}
\label{sec:related_bbc}

Empirical studies measure how often behavioral breaking changes occur and how they reach the downstream project. Mostafa et al.\ inspect 126 behavioral backward incompatibilities in 15 Java libraries~\cite{10.1145/3092703.3092721}. Kong et al.\ find that behavior changes are the largest category of breaking changes marked by npm maintainers, at 68.1\%~\cite{10.1145/3702991}. Jayasuriya et al.\ find that behavioral breaking changes reach client test suites as exceptions or assertion failures~\cite{10.1145/3643782}. Montandon et al.\ find that changed default argument values in Scikit-Learn affect 35\% of the clients they analyze~\cite{10.1016/j.jss.2025.112460}.

Detection tools report the changed APIs. Static tools compare the two library versions without executing them. Sembid compares the semantic graphs of APIs whose signatures are unchanged and, as its authors report, misses changes that are only triggered dynamically~\cite{10.1145/3551349.3556956}, UnCheckGuard lists the unchecked exceptions that a new version adds~\cite{11190106}, and Uppdatera reports the changed functions that the client reaches in a static call graph~\cite{HEJDERUP2022111097}. Dynamic tools execute code to observe the change. DeBBI runs the test suites of client projects against the new version~\cite{9283994}, and GILESI, the closest to differential execution, records the inputs and outputs of every library call during client test execution and compares them across versions~\cite{monce2025gilesi}. SENSOR and CompCheck generate tests that reveal semantic differences between two library versions~\cite{9350237, 10.1145/3582569}, and Testora does the same for the two sides of one pull request~\cite{10.1145/3744916.3764527}. All of these tools stop at detection, whereas \approach{} filters the library diff with the first divergence and gives both items to the repair agent.

\subsection{Repairing Breaking Changes}
\label{sec:related_repair}

Before LLMs, repair methods apply transformation rules from different sources. SemDiff recommends the adaptation that the framework makes to its own code~\cite{4814159}, APIfix synthesizes rules from adaptations of other clients~\cite{10.1145/3485538}, MELT mines rules from merged library pull requests~\cite{10.1109/ASE56229.2023.00117}, and JSFIX applies semantic patches written for the change~\cite{9402106}. All four need an existing adaptation or a written description, and a behavioral change that no one has adapted or described gives neither.

LLM-based methods form three lines by their source of evidence. The first line starts from the compile error of the broken build. Byam rewrites each failing file in one LLM call with the interface diff of the upgrade in the prompt~\cite{10.1007/s10664-026-10835-1}, Fruntke and Krinke let an agent read the repository and revise patches on build feedback~\cite{10.1145/3729366}, BigBag generates a reusable AST transformation program~\cite{reyes2026agenticgenerationasttransformation}, and DependaFix packages Byam as a GitHub App~\cite{10.1145/3803437.3806404}. All four handle only updates that fail at compilation, and Byam reports that semantic failures appear once compilation is restored. The second line repairs from what the maintainers of the library write. Tawosi et al.\ drive a multi-agent framework with migration documentation~\cite{tawosi2025llmagentsautomateddependency}, BDUpdater, which we replicate in our evaluation, mines breaking change records from the upstream repository~\cite{10.1145/3808116}, and DepRepair distills release notes into a rule table~\cite{yang2026deprepairllmbasedsourcecoderepair}. The third line reads the library diff. AIMigrate prompts the LLM once per project file with the diff between the two library versions~\cite{rosenfeld2025diffmakesautomatingcode}. The tool does not run the project, so the diff is neither cut down to the break nor checked against it. Benchmarks accompany these methods. Section~\ref{sec:benchmark} evaluates BUMP~\cite{10589737}, CompSuite~\cite{10.1109/ASE56229.2023.00127}, TimeMachine-bench~\cite{fujii-etal-2026-timemachine}, and RepoRescue~\cite{lin2026reporescueempiricalstudyllm} against the requirements of the repair task. MigrationBench~\cite{10.1145/3770855.3817526} and FreshBrew~\cite{may2025freshbrewbenchmarkevaluatingai} target migration away from Java 8, and GitChameleon 2.0 poses version-conditioned Python completion problems~\cite{misra2025gitchameleon20evaluatingai}, so all three test whether the model writes code for a given version, not whether it repairs a broken downstream project. In sum, prior methods start from a compile error, a written description, or an unfiltered library diff, whereas \approach{} starts from the behavior change observed by running the failing test under both versions.

\section{Conclusion}
\label{sec:conclusion}

This paper finds that existing LLM-based repair methods and Pure Agent often fail to obtain the upstream evidence that a successful and efficient repair of a behavioral breaking change requires. Based on this finding, we propose \approach{}, a repair method that runs the failing test under the old and new library versions, ranks the calls whose return value differs to identify a candidate root API, and filters the library diff with that candidate. The paper also contributes \benchmark{}, a benchmark of 100 behavioral dependency breaks in Python and JavaScript. On \benchmark{}, \approach{} raises the average pass rate by up to 17\% relative to Pure Agent and needs up to 27\% fewer agent steps per successful repair. These results show that \approach{} repairs behavioral breaking changes more effectively and efficiently.

\section{Data Availability}
\label{sec:data}

The replication package contains the implementation of \approach{} and \benchmark{}, so the results can be reproduced directly. The package is available at \url{https://github.com/Elendil3703/BBCFixer}.

\bibliographystyle{ACM-Reference-Format}
\bibliography{references}


\end{document}